\documentstyle[11pt,a4]{article}

\def \square {\hbox{$\sqcup\!\!\!\!\sqcap$}} 
\newcommand{\be}{\begin{equation}}
\newcommand{\ee}{\end{equation}} 
\newcommand{\bea}{\begin{eqnarray}}
\newcommand{\eea}{\end{eqnarray}}

\begin{document}

\begin{titlepage}

\bigskip

\begin{center}

{\bf \LARGE Free Fields for Chiral 2D Dilaton  
Gravity. 	}
\footnote{Work partially supported by the 
{\it Comisi\'on Interministerial de Ciencia y Tecnolog\'{\i}a}\/ 
and {\it DGICYT}} 

\bigskip 

 J. Cruz$^a$\footnote{\sc cruz@lie.uv.es},
 J. Navarro-Salas$^a$\footnote{\sc jnavarro@lie.uv.es} and
M. Navarro$^b$.\footnote{\sc mnavarro@fismat2.ugr.es}

\end{center}

\bigskip%

\footnotesize
a) Departamento de F\'{\i}sica Te\'orica and 
	IFIC, Centro Mixto Universidad de Valencia-CSIC.
	Facultad de F\'{\i}sica, Universidad de Valencia,	
        Burjassot-46100, Valencia, Spain. 
  \newline
 
  b)       Instituto Carlos I de F\'\i sica Te\'orica y Computacional,
        Facultad  de  Ciencias, Universidad de Granada. 
        Campus de Fuentenueva, 18002, Granada, Spain.      
\normalsize 

\bigskip
\bigskip


\begin{center}
			{\bf Abstract}
\end{center}
We give an
  explicit canonical transformation which transforms a
  generic chiral 2D dilaton gravity
 model into a free field theory.
 \newline
 \newline
 PACS number(s): 04.60.Kz, 04.60.Ds
 \newline
 \end{titlepage}
 \newpage
 In the last years, a great effort has been dedicated to study two-dimensional gravity theories.
 They provide a simplified setting to study quantum aspects of four-dimensional gravity. 
 The simplest one of these models was introduced by Callan, Giddigs, Harvey and Strominger (CGHS)
 This theory can be canonically mapped into a theory 
 of two free fields with a Minkowskian target space \cite{multiple}.
 In the new variables it is possible to carry out a consistent functional Schr\"odinger
 quantization of the matter-coupled CGHS theory
 \cite{Jackiw, Kuchar}. This result has been recently extended to the Jackiw-Teitelboim model and
  the model with an exponential potential by finding explicit canonical
   transformations which also convert these theories
 into a theory of two free fields with a flat target space of Minkowskian signature
\cite{Cruz1}.
 In fact, it has been proven that such a transformation exists for an arbitrary
 model of 2D dilaton gravity 
 \cite{Cruz2}
 although the explicit form remains elusive except for the above mentioned cases.
 In this paper we will give an explicit expression for the 
 canonical transformation which maps a generic 2D dilaton gravity model into a 
 free field theory
 when chiral matter is coupled.
 
It is well-known that the action of a generic 2D dilaton gravity model can be brought to the form
\cite{Banks}
 \be
 S=\int d^2 x\sqrt{-g}\left[R\phi+V\left(\phi\right)+{1\over2}\left(\nabla f\right)^2\right]
 \>,\label{i}
 \ee
 by a conformal reparametrization of the fields.
 The equations of motion derived from (\ref{i}) are
 \be
 R+V^{\prime}\left(\phi\right)=0
 \>,\label{ii}
 \ee
 \be
 \nabla_{\mu}\nabla_{\nu}\phi-{1\over2}g_{\mu\nu}V\left(\phi\right)=
 {1\over2}T^f_{\mu\nu}={1\over2}\left(\nabla_{\mu}f\nabla_{\nu}f-{1\over2}g_{\mu\nu}\left(\nabla f\right)^2\right)
 \>,\label{iii}
 \ee
 \be
 \square f=0
 \>.\label{iv}
 \ee
 This theory is equivalent, via a canonical transformation, to a theory of three free fields.
 However, as we have already mentioned the explicit form of the field transformation 
 is unknown, up to same particular cases.
 If we restrict the phase space of the theory by impossing the condition
 $T^f_{\mu\nu}T^{f\mu\nu}=0$, which physically means to restrict the matter degrees 
 to freedom to (let say)
 left-movers, the
 canonical transformation relating dilaton-gravity and free fields variables
 of the modified (chiral) theory
 can be explicitly given in the general case.
 A convenient form for the metric, which is useful 
 to study the chiral matter-coupled theory, is the following
 \cite{Bardeen}
 \be
 ds^2=e^{2\rho}Adv^2+2e^{\rho}dvd\phi
\>,\label{v}
\ee
where $\phi$ is the dilaton field. In this way the dilaton is taken to
 be a coordinate and the two degrees of freedom are $\rho$ and $A$.
 The chirality condition is
 automatically satisfied if we take $T^f_{vv}= T^f_{vv}\left(v\right)$ as the only non-vanishing
 component of the energy momentum tensor.
 It is easy to check that with this condition
  the covariant conservation of the energy momentum tensor 
 (i.e, the equation of motion for the matter field) holds identically.
 The different components of  the equation (\ref{iii})
 are
 \be
 \rho_{,\phi}=-{1\over2}T^f_{\phi\phi}=0
 \>,\label{vi}
 \ee
 \be
 -e^{\rho}A_{,v}=T^f_{vv}={1\over2}f_{,v}^2
 \>,\label{vii}
 \ee
 \be
 A_{,\phi}+V\left(\phi\right)=-A\rho_{,\phi}
 \>.\label{viii}
 \ee
 From equation (\ref{vi}) we see that $\rho$ is a chiral field $\rho=\rho\left(v\right)$ while 
 equation (\ref{viii}) can be written as
 \be
 \left(A+J\left(\phi\right)\right)_{,\phi}=0
 \>,\label{ix}
 \ee
 where ${dJ\left(\phi\right)\over d\phi}=V\left(\phi\right)$. So, the solution can be expressed as 
 \be
 A+J\left(\phi\right)=-2E
 \>,\label{x}
 \ee
 with $E=E\left(v\right)$ an arbitrary chiral function.
 This function turns out to be the local energy \cite{Gegenberg}
 \be
 E={1\over2}\left[\left(\nabla\phi\right)^2-J\left(\phi\right)\right]
 \>,\label{xi}
 \ee
 which is a conserved quantity when there is no matter at all.
 The constraint equation (\ref{vii}) becomes now
 \be
 2e^{\rho}E_{,v}=T^f_{vv}
 \>.\label{xii}
 \ee
 Therefore the general solution can be written as
 \be
 ds^2=-e^{2\rho}\left(\int e^{-\rho}T^f_{vv}dv+J\left(\phi\right)\right)
 dv^2+2e^{\rho}d\phi dv
 \>,\label{xiii}
 \ee
 with $\rho=\rho\left(v\right)$ an arbitrary function which is 
 associated with reparametrizations of the coordinate $v$.
If we choose $\rho=0$ the above expression is a generalization of the 
Vaidya solution of spherically symmetric gravitational collapse
for an arbitrary 2D dilaton gravity model.
Our aim now is to employ this result to construct a canonical transformation 
which maps the theory into a free  
 field theory. 
 To this end we shall work out the symplectic 2-form on the phase-space  
 constrained by the chirality condition. A useful way to achieve this is by using the 
 covariant phase-space
 formalism \cite{Witten} which automatically incorporates the constraint
 in a consistent way. The symplectic form of the theory is given by
 \be
 \omega=\delta\int_{\Sigma} j^{\mu}d\sigma_{\mu}
 \>,\ee
 where $\delta$ stands for the exterior differential on phase space and $j^{\mu}$
 is given by
 \be
 j^{\mu}=-\phi(g^{\alpha\beta}\nabla^{\mu}\delta g_{\alpha\beta}-g^{\mu\alpha}
 \nabla^{\beta}
 \delta g_{\alpha\beta})+\nabla^{\mu}\phi g_{\alpha\beta}\delta g^{\alpha\beta}
 -\nabla_{\alpha}\phi\delta g^{\alpha\mu}+\nabla^{\mu}f\delta f
 \>.
 \ee
 To evaluate $\omega$ on the space of chiral solutions it is convenient
 to choose $\phi=constant$ as the initial data surface. The result is
 \be
 \omega=\int dv \left[\delta(-2E)\delta e^{\rho}+\delta f\delta f_{,v}\right]
 \>,\ee

implying the following Poisson bracket of dilaton-gravity variables
 \be
 \left\{-2E\left(v\right), e^{\rho\left(\tilde v\right)}\right\}
 =\delta\left(v-\tilde v\right)
 \>.\label{xv}
 \ee
This is a generalization of the reassoning used in Ref\cite{Verlinde2}
for the particular case of the CGHS theory.

The phase-space of the chiral theory is, therefore, made out of two arbitrary
(chiral) functions ($E\left(v\right),\rho\left(v\right)$) in addition to the left-mover
matter field $f(v)$.
Defining
 \be
 X=-2E\>,\label{xvi}
 \ee
 \be
 \Pi=e^{\rho}\>,\label{xvii}
 \ee
 the constraint (\ref{xii}) becomes the constraint of a free field theory
 \be
 T^f_{vv}+\Pi X_{,v}=0
 \>.\label{xviii}
 \ee
 The above formulas are the main result of this paper and provide the explicit realization of the 
  canonical transformation proposed in Ref\cite{Cruz2} for the chiral matter case.
 To finish this paper it is interesting to comment that in the absence of matter
  $\left(T^f_{vv}=0\right)$, 
 $E$ is a constant of motion and the Poisson bracket (\ref{xv})
 converts into 
 \be
 \left\{-2E,\int dv \ e^{\rho}\right\}=1
 \>,\label{xix}  
  \ee
 indicating that the only (global) degree of freedom is 
 $C=-2E$ and the canonically conjugated variable is $P=\int dv\ e^{\rho}$ in agreement with
 \cite{Louis}.

 \section*{Acknowledgements}
 J. C. acknowledges the Generalitat Valenciana for a FPI grant. 
 M. N. acknowledges the Spanish MEC, CSIC and IMAFF (Madrid) for a research contract.

 \end{document}